# γ Doradus variables, a new class of pulsating stars. The case of HD 224945

E. Poretti[1], C. Akan[2], M. Bossi[1], C. Koen[3], K.Krisciunas[4], E. Rodriguez[5]

[1] Osservatorio Astronomico di Brera, Merate, Italy
[2] Ege University, Turkey
[3] South African Astronomical Observatory, Cape Town, South Africa
[4] Joint Astronomy Center, Hawaii, USA
[5] Instituto de Astronomia y Astrofisica, Granada, Spain

## ABSTRACT

A multisite campaign involving five observatories allowed us to solve the light curve of the γ Doradus star HD 224945. The multiperiodic content, the high frequency term at 3.00 c/d and the frequency spacing between the observed terms strongly support its pulsational nature. Considering the low frequency terms observed in γ Dor stars, it seems that $g$–mode pulsators can be found near the cold border of the instability strip.

## 1. The γ Doradus variables

In recent years a few early F–type stars showing small amplitude light variations (a few hundredths of a mag) have been discovered (for a reference list see Krisciunas & Handler 1995). These variations are not regular, but their analysis suggested that they can result from the superposition of close periodicities with characteristic time scales of the order of 1 day. Since these stars lie on and beyond the cold border of the instability strip, it was not an easy task to establish the physical cause of the variability, detected on the basis of fragmentary, often single–site, observations. It should be noted that several of the new variables were first selected as comparison stars for other targets precisely because their location in the HR diagram did not suggest any variability. In all the papers announcing a new member of the class, rotational modulation of features on the surface or $g$–mode pulsation were discussed as the possible physical mechanisms of variability. At the occasion of the IAU Colloquium 155 (*Astrophysical Applications of Stellar Pulsation*, Cape Town, February 1995), these stars were named "γ Doradus variables" after the brightest member; moreover, a general consensus on their pulsational nature was achieved, but it was also recognized that an observational effort should be made to obtain clearer evidence.

Researchers of four institutes were involved in the scientific discussion and in planning new observational studies on their own preferred targets: K. Krisciunas (JAC Hawaii, 9 Aur), L. Balona (South African Astronomical Observatory, γ Dor), Mantegazza and Poretti (Brera Observatory, HD 224945 and HD 224638), F. Zerbi (on leave from Pavia University) and E. Rodriguez (Sierra Nevada Observatory, HD 164615). The new observing campaigns were structured in a multisite way, in order to overcome the effect of the 1 c/d alias problem, of paramount importance here. As regards HD 224945 and HD 224638, we remind the reader that even dealing with very accurate measurements carried out at ESO (La Silla, Chile), Mantegazza, Poretti & Zerbi (1994; hereinafter Paper I) could not obtain a clear picture of their photometric behaviour since two possibilities remained open to explain their complicated light curves:

*i)* one or two periodicities having double– or triple–wave light curves, suggesting rotational modulation;

*ii)* a multifrequency content, i.e. a number of independent modes simultaneously excited.

All the people involved participated enthusiastically in the campaigns: Balona et al. (1996) have already reported their results on γ Dor, Zerbi et al. (1996) on 9 Aur; other data are under analysis. Here we present the results obtained on HD 224945.

## 2. The observations and the data reduction

A campaign was organized in October 1995 to monitor HD 224945 and HD 224638. Five sites were involved: ESO (La Silla, Chile, observer E. Poretti), Ege Observatory (Turkey, observer: C. Akan), South Africa (SAAO, observer C. Koen), Sierra Nevada (OSN, observers E. Rodriguez and S. Martin), Hawaii (Mauna Kea Observatory, observers K. Krisciunas, R. Crowe and M. Roberts). It should be noted that owing to their equatorial positions, the two stars were visible from both the hemispheres. At all the sites, weather conditions were not the best possible: at ESO the sky was usually photometric (standard deviation of the check star measurements: 2.9 mmag), but 4 (of 12) nights were cloudly, at Hawaii and SAAO photometric conditions were not always satisfied (s.d. 5.1 and 5.5 mag, respectively), at Sierra Nevada only the first week of the campaign found clear skies (s.d. 3.2 mmag), at Ege Observatory observing conditions were a bit worse than expected (s.d. 10.2 mmag). The raw data was first scrutinized by the observers themselves, but the final differential magnitudes were calculated by the same code, i.e. we avoid different data reduction routines. In spite of this, small misalignments were observed (due to small differences in the $\lambda_{eq}$ values), but the good longitude distribution allowed us to calculate systematic differences by using measurement overlaps (Turkey and South Africa, South Africa and Spain, Spain and Chile, Chile and Hawaii). Hence, we obtained homogenous sets of data with a rather clear spectral window (Fig. 1).

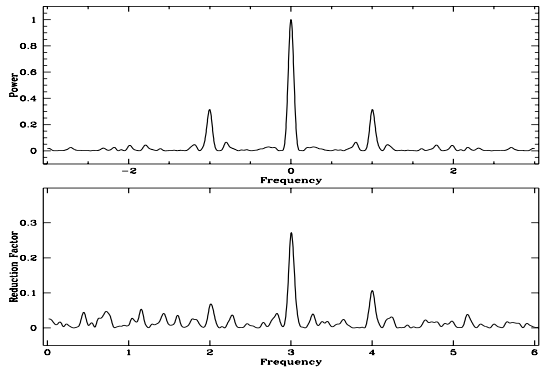

Figure 1: Top: Spectral window of the 1995 multisite campaign. Bottom: Power spectrum of the 1995 data of HD224945

### 3. The frequency analysis

Photometry was carried out in $B$ and $V$ passbands, but to allow us a comparison with the older datasets we discuss only the $B$ data. The analysis of the HD 224945 data was of particular interest since the results of Paper I indicated that this star shows the highest frequency terms in the $\gamma$ Dor class. Hence, HD 224945 could be of decisive importance in the scientific debate. The least–squares power spectrum (lower panel of Fig. 1) clearly shows the highest peak at 3.00 c/d. A number of tests showed that this term cannot be attributed to an instrumental or atmospheric effect. In Paper I a 2.00 c/d term was considered in the discussion, but it was remarked that this identification suffered from a 1 c/d alias effect. Further analysis of the 1995 data allowed us to detect unambiguously two other terms (2.83 and 1.16 c/d). We then revisited the 1991 data, finding differences more substantial than the 2.00/3.00 c/d interchange:

1. In 1995 the 3.00 c/d term had by far the largest amplitude (as can be deduced from Fig. 1 and as is shown in Table 1), while in 1991 the detected terms had a similar amplitude (compare also the lower panel of Fig. 1 with the upper panel of Fig. 6 in Paper I);

2. The 2.83 c/d term is common between the 1991 and 1995 datasets, but the 1991 amplitude is twice the 1995 one;

3. The 1.16 c/d term was not detected in the 1991 data; in turn, the 2.31 c/d term, easily detected in 1991 data, is not clearly discernible in 1995 data. Since the frequency ratio is very close to 2:1, the amplitude changes (as a result of resonance, switching effects ...) deserve further investigation.

Table 1: A comparison between the photometric solutions of the three observing seasons of HD 224945

| | Amplitude [mag] | | |
|---|---|---|---|
| Frequency | 1995 | 1994 | 1991 |
| 3.0015 | 0.0060 | 0.0041 | 0.0041 |
| 2.8331 | 0.0023 | 0.0041 | 0.0046 |
| 1.1603 | 0.0031 | 0.0025 | – |
| 2.3107 | – | 0.0026 | 0.0043 |

### 4. Conclusion

Table 1 supplies a comparison between the least–squares fits of the 1991 and 1995 data; moreover, the same technique was also applied to the data collected by M. Bossi in 1994 at ESO. The three terms detected unequivocally in the 1995 data (free from the 1 c/d alias problem) can hardly be reconciled with the double– and triple–waves suggested on the basis of the 1991 data. It is definitely established that the light curve of HD 224945 has a multiperiodic content; this object is the $\gamma$ Doradus variable star showing the highest frequency and the largest separation between the three dominant modes (3.00, 2.83, 1.16 c/d). These two observational facts rule out the possibility of a modulation with the rotation period and is strong evidence for a pulsation. Moreover, comparison of the results for three different seasons suggests amplitude variations.